\documentclass[a4paper,11pt]{article}
\usepackage{amstext}
\usepackage{amsgen}
\usepackage{amssymb}
\usepackage{amsfonts}
\usepackage{latexsym}
\usepackage{amsmath}
\usepackage{theorem}
\usepackage{pifont}
\usepackage{txfonts}

 \newcommand{\hs}[1]{\hspace*{ #1 mm}}

 \newcommand{\integer}{\mathbb{Z}}

 \newcommand{\complex}{\mathbb{C}}

 \newcommand{\HH}{{\cal H}}
 
 \newcommand{\KK}{{\cal K}}

\theoremstyle{plain}
\theorembodyfont{\rmfamily}
 \newtheorem{theorem}{Theorem}[section]
 \newtheorem{lemma}[theorem]{Lemma}
 
 \newtheorem{corollary}[theorem]{Corollary}

 {\theorembodyfont{\rmfamily}
\newtheorem{definition}[theorem]{Definition}}
 {\theorembodyfont{\rmfamily} }
 {\theorembodyfont{\rmfamily} }
 {\theorembodyfont{\rmfamily} }

 \newenvironment{proof}{\par \noindent
            {\bf Proof. \hs{2}}}{\hfill$\Box$ \vspace*{3mm}}

 \newcommand{\bra}[1]{\langle #1 |}
 \newcommand{\ket}[1]{| #1 \rangle}
 
 \newcommand{\ketbra}[2]{| #1 \rangle\langle #2 |}

 \newcommand{\tr}{\mathrm{tr}}
 \newcommand{\rank}{\mathrm{rank}}
 \newcommand{\supp}{\mathrm{supp}}
 \newcommand{\norm}[1]{\Vert #1 \Vert}
 \newcommand{\Norm}[1]{\left\Vert #1 \right\Vert}
 
 \newcommand{\tracenorm}[1]{\| #1 \|_\mathrm{tr}}
 \newcommand{\Tracenorm}[1]{\left\| #1 \right\|_\mathrm{tr}}

 \def\magicwand{\itemsep=0pt\parskip=0pt\topskip=0pt}

\newif\ifnotesw\noteswtrue
   {\ifnotesw\marginpar[\hfill\(\top\)]{\(\top\)}\fi}%
      {\ifnotesw\marginpar[\hfill\(\bot\)]{\(\bot\)}\fi}
      
\newcommand{\mnote}[1]%
   {\ifnotesw\marginpar%
	  [{\scriptsize\begin{minipage}[t]{\marginparwidth}
	  \raggedleft#1%
		  \end{minipage}}]%
	  {\scriptsize\begin{minipage}[t]{\marginparwidth}
	  \raggedright#1%
		  \end{minipage}}%
    \fi}

\newcommand{\ignore}[1]{}

\renewcommand{\HH}{\mathcal{H}}

\renewcommand{\KK}{\mathcal{K}}

\begin{document}

\sloppy

\title{\LARGE
  \textbf{
    Quantum Measurements for Hidden Subgroup Problems with Optimal Sample Complexity
  }
}

\author{
  $\text{Masahito Hayashi}^{\ast}$\\
  \texttt{masahito@qci.jst.go.jp}
  \and
  $\text{Akinori Kawachi}^{\dagger}$\\
  \texttt{kawachi@is.titech.ac.jp}
  \and
  $\text{Hirotada Kobayashi}^{\ddagger}$\\
  \texttt{hirotada@nii.ac.jp}
}

\date{}

\maketitle
\thispagestyle{plain}
\pagestyle{plain}

\begin{center}
{\large
  ${}^{\ast}$ERATO-SORST Quantum Computation and Information Project\\
  Japan Science and Technology Agency\\
  5-28-3 Hongo, Bunkyo-ku, Tokyo 113-0033, Japan\\
[3mm]
  ${}^{\dagger}$Department of Mathematical and Computing Sciences\\
  Tokyo Institute of Technology\\
  2-12-1 Ookayama, Meguro-ku, Tokyo 152-8552, Japan\\
[3mm]
  ${}^{\ddagger}$Principles of Informatics Research Division\\
  National Institute of Informatics\\
  2-1-2 Hitotsubashi, Chiyoda-ku, Tokyo 101-8430, Japan
}\\
\end{center}

\begin{abstract}
One of the central issues in the hidden subgroup problem is to bound
the sample complexity, i.e., the number of identical samples of coset states sufficient and necessary to solve the problem. In this paper, we present general bounds for the sample complexity of the identification and decision versions of the hidden subgroup problem. As a consequence of the bounds, we show that the sample complexity for both of the decision and identification versions is $\Theta(\log|\HH|/\log p)$ for a candidate set $\HH$ of hidden subgroups in the case that the candidate subgroups have the same prime order $p$, which implies that the decision version is at least as hard as the identification version in this case. In particular, it does so for the important instances such as the dihedral and the symmetric hidden subgroup problems. Moreover, the upper bound of the identification is attained by the pretty good measurement. This shows that the pretty good measurement can identify any hidden subgroup of an arbitrary group with at most $O(\log|\HH|)$ samples.
\end{abstract}

\section{Introduction}
\subsection{Background}
The {\em hidden subgroup problem\/} is one of the central issues in 
quantum computation, which was introduced for revealing the structure 
behind exponential speedups in quantum computation \cite{ME99}.
\begin{definition}[Hidden Subgroup Problem (HSP)]
Let $G$ be a finite group.
For a hidden subgroup $H\leq G$, we define a map $f_H$ 
from $G$ to a finite set $S$ with the property that 
$f_H(g)=f_H(gh)$ if and only if $h\in H$.
Given $f_H: G\rightarrow S$ and a generator set of $G$,
the hidden subgroup problem (HSP) is the 
problem of finding a set of generators for the hidden subgroup $H$.
We say that HSP over $G$ is efficiently solvable if 
we can construct an algorithm in time polynomial in $\log{|G|}$.
\end{definition}
The nature of many existing quantum algorithms relies on 
efficient solutions to Abelian HSPs (i.e., HSPs over Abelian groups) 
\cite{Sim97,Kit95,BL95,BH97}.
In particular, Shor's cerebrated quantum algorithms for factoring and 
discrete logarithm essentially consist of reductions to 
certain Abelian HSPs and efficient solutions to the Abelian HSPs \cite{Sho97}. 
Besides his results, many efficient quantum algorithms 
for important number-theoretic problems (e.g., Pell's equation \cite{Hal02} 
and unit group of a number field \cite{Hal05,SV05}) were based on 
solutions to Abelian HSPs.

Recently, non-Abelian HSPs have also received much attention.
It is well known that the graph isomorphism problem can be reduced to the
HSP over the symmetric group \cite{BL95,Bea97} 
(more strictly, the HSP over $S_n\wr S_2$ \cite{EHb99}).
Regev showed that we can construct an efficient quantum algorithm 
for the unique shortest vector problem if we find an efficient solution to HSP 
over the dihedral group under certain conditions \cite{Reg04a}.
While the efficient quantum algorithm for general Abelian HSPs has been already
given \cite{Kit95,ME99}, 
the non-Abelian HSPs are extremely harder than the Abelian ones.
There actually exist efficient quantum algorithms for HSPs over 
several special classes of non-Abelian groups \cite{RB98,FIMSS03,HRT03,Gav04,GSVV04,IL04,IMS03,MRRS04,BCD05b}.
Nonetheless, most of important cases of non-Abelian HSPs, including 
the dihedral and symmetric HSPs, are not known to have efficient solutions.
Thus, finding efficient algorithms for non-Abelian HSPs is one of the most 
challenging issues in quantum computation.

The main approach to the non-Abelian HSPs is based on a generic framework 
called the {\em standard method}. To our best knowledge, 
all the existing quantum algorithms for HSPs essentially contain this 
framework.
The standard method essentially reduces HSPs to the quantum state identification\cite{Sen06} for the so-called {\em coset states}, which contain information of the hidden subgroup.
\begin{definition}[Coset State and Standard Method]
Let $G$ be any finite group and $H$ be
the hidden subgroup of $G$. We then define the 
{\em coset state} $\rho_{H}$ for $H$ as
$
 \rho_{H} 
= \frac{1}{|G|}\sum_{g\in G}\ketbra{gH}{gH}
= \frac{|H|}{|G|}\sum_{g\in G/H}\ketbra{gH}{gH},
$
where
$
 \ket{gH} = \frac{1}{\sqrt{|H|}}\sum_{h\in H}\ket{gh}.
$
\begin{description}
\magicwand
\item[Standard Method with $\boldsymbol{k}$ Coset States]
\item[{\rm (1)}] Prepare two registers with a uniform superposition over $G$ in the first register and all zeros in the second register: $\frac{1}{\sqrt{|G|}}\sum_{g\in G}\ket{g}\ket{0}$.
\item[{\rm (2)}] Compute $f_H(g)$ and store the result to the second register: 
$\frac{1}{\sqrt{|G|}}\sum_{g\in G}\ket{g}\ket{f_H(g)}$.
\item[{\rm (3)}] Discard the second register to obtain a coset state: $\rho_H=\frac{|H|}{|G|}\sum_{g\in G/H}\ketbra{gH}{gH}$.
\item[{\rm (4)}] Repeat (1)--(3) $k$ times and then apply a quantum 
measurement to $k$ samples of $\rho_H$.
\end{description}
\end{definition}
Thus the main task for solving HSP based on the standard method
is to find an efficiently implementable quantum measurement extracting 
the information of the hidden subgroup from identical samples of the coset state.

Many researchers have broadly studied hard instances of non-Abelian HSPs 
from positive and negative aspects based on the standard method. 
In particular, they have focused on the {\em sample complexity\/} of HSPs, 
i.e., how many coset states are sufficient and necessary 
to identify the hidden subgroup with a constant success probability.

In several classes of the non-Abelian HSPs 
for which efficient algorithms are unknown, 
it is shown that we can identify any hidden subgroup by 
(possibly inefficient) classical post-processes using the classical 
information obtained by the quantum Fourier transforms to polynomially 
many samples of coset states \cite{EH00,HRT03,GSVV04,MRRS04}.

Bacon, Childs and van Dam demonstrated that the so-called {\em pretty good 
measurement\/} (PGM, also known as the {\em squire root measurement\/} or 
{\em least squares measurement\/} \cite{HW94}) 
is optimal for identifying coset states
in view of the sample complexity
on a class of semidirect product groups $A\rtimes\integer_p$ including the dihedral group, where $A$ is any Abelian group 
and $p$ is a prime \cite{BCD05b}.
They proved that the sample complexity is $\Theta(\log{|A|}/\log{p})$ 
to identify the hidden subgroup by the PGM from the candidate set $\HH_\mathrm{SDP}=\{\langle(a,1)\rangle < A\rtimes \integer_p : a\in A \}$.
Moore and Russell generalized their result to prove the optimality of 
the PGM for a wider class of HSPs \cite{MR05d}.
They actually gave the PGM for identifying 
coset states of hidden conjugates of a subgroup, i.e., hidden subgroups 
having form of $g^{-1}H g$ for a fixed non-normal subgroup $H$ of a finite
group $G$ and $g\in G$. These results of \cite{BCD05b,MR05d} showed 
that the PGM succeeds for a wide class of HSPs with 
at most $O(\log|\HH|)$ samples for the candidate set $\HH$ of hidden subgroups.
For a more general case, Ettinger, H{\o}yer and Knill gave a 
bounded-error quantum measurement that solves HSP over any finite group $G$ with $O(\log^2|G|)$ samples of coset states (Theorem 2 in \cite{EHK04}).
They also constructed an error-free measurement for the general HSP
with the same sample complexity $O(\log^2|G|)$ within a constant 
factor in \cite{EHK04} by combining the bounded-error one with 
the amplitude amplification technique \cite{BHMT02}.

These quantum measurements ignore the time complexity issue in 
general. However, they may lead to efficient quantum algorithms for HSPs. 
Bacon et al.~actually gave efficient implementation of the PGM for 
identifying given coset states on a class of the semidirect groups including the Heisenberg group \cite{BCD05b}, i.e., 
they constructed an efficient quantum algorithm for the HSPs from 
the corresponding PGMs.
Hence, to give the quantum measurements for identification of given coset 
states like PGMs may play important roles towards the construction of 
efficient quantum algorithms for HSPs.

The negative results of the standard method has also been studied from 
an information-theoretic viewpoint, which are based on a decision version 
of the HSP defined as the problem of deciding whether the hidden subgroup 
is trivial or not. In particular, the difficulty of the HSP over the symmetric 
group $S_n$ has been shown by a number of results for this decision 
version \cite{HRT03,GSVV04,KS05,MRS05,MR05a}. Hallgren et al.~recently proved 
that a joint  measurement across multiple samples of coset states is 
essentially required to solve a decision version over the symmetric group, 
which is deeply related to the graph isomorphism problem. More precisely, 
they showed that joint quantum measurements across $\Omega(n\log{n})$ samples 
of coset states are necessary to decide whether the given samples are 
generated from the trivial subgroup $\{id\}$ or a subgroup 
in $\HH_\mathrm{Sym}=\{ H< S_n : H=\langle h\rangle,\ h^2=id,\ h(i)\neq i
\ (i=1,...,n)\}$, i.e., a set of all the subgroups generated by 
the involution composed of $n/2$ disjoint transpositions \cite{HMRRS06}.

\subsection{Our Contributions}
We study upper and lower bounds for the sample complexity of general HSPs from an information-theoretic viewpoint. We consider two problems associated with HSPs to deal with their sample complexity. The first one is the identification version for solving HSPs based on the standard method.
\begin{definition}[Coset State Identification (CSI)]
Let $\HH$ be a set of candidate subgroups of a finite group $G$.
We then define $S_\HH$ as a set of coset states corresponding to $\HH$.
Given a black box that generates an unknown coset state $\rho_H$ in $S_\HH$,
the Coset State Identification (CSI) for $\HH$ is the problem of identifying 
$H\in\HH$.
\end{definition}
One can easily see that any solution to HSP based on the standard method 
reduces this identification of coset states.
We now define the sample complexity of CSI for $\HH$ as the sufficient 
and necessary number of samples for identifying the given coset state 
with a constant probability.

The second one is the decision version, named the Triviality of Coset State.
Special cases of this problem have been discussed 
for the limitations of the standard method in many previous results \cite{HRT03,GSVV04,KS05,MR05a,MRS05,AMR05,HMRRS06}.
\begin{definition}[Triviality of Coset State (TCS)]\label{TCS}
Let $\HH$ be a set of candidate non-trivial subgroups of a finite 
group $G$, i.e., $H \neq \{id\}$ for every $H\in\HH$.
We then define $S_\HH$ as a set of coset states
corresponding to $\HH$.
Given a black box that generates an unknown state $\sigma$ 
that is either in $S_\HH$ (i.e., a coset state for the non-trivial subgroup) 
or equal to $I/|G|$ (i.e., a coset state for the trivial subgroup), 
the Triviality of Coset State for $S_\HH$ is the problem of deciding 
whether $\sigma$ is in $S_\HH$ or equal to $I/|G|$.
We say that a quantum algorithm solves TCS with a constant advantage
 if it correctly decides whether a given state is in $S_\HH$ or 
equal to $I/|G|$ with success probability at least $1/2+\delta$ for 
some constant $\delta\in(0,1/2]$.
\end{definition}
Similarly to the case of CSI, we define the sample complexity of TCS for 
$\HH$ as the sufficient and necessary number of coset states to 
solve TCS with a constant advantage.

Note that this problem might be efficiently solvable even if we cannot 
identify the hidden subgroup. Actually, if we can give a solution to TCS for 
$\HH_\mathrm{Sym}=\{ H< S_n : H=\langle h\rangle,\ h^2=id,\ h(i)\neq i
\ (i=1,...,n)\}$, we can also solve the rigid graph isomorphism problem, i.e., 
the problem of finding an 
isomorphism between two graphs having no non-trivial automorphisms, and 
the decisional graph automorphism problem, i.e., the problem of deciding 
whether a given graph has non-trivial automorphisms or not \cite{KST93}.

In this paper, we give bounds of the sample complexity of CSI and TCS 
by simple information-theoretic arguments. We present the following bounds
of the sample complexity of CSI.
\begin{theorem}[Upper and Lower Bounds for CSI]\label{ULBforCSI}
Let $\HH$ be any set of candidate subgroups of a finite group.
Then, the sample complexity of CSI for $\HH$
is at most $O\left(\frac{\log|\HH|}{\log{{\min_{H\neq H'\in\HH}(|H|/|H\cap H'|)}}}\right)$ and at least $\Omega\left(\frac{\log|\HH|}{\log{\max_{H\in\HH}|H|}}\right)$.
\end{theorem}
Moreover, the upper bound of CSI can be attained by the PGM. 
This shows that we can identify a hidden subgroup for an arbitrary 
group $G$ by the PGM with at most $O(\log|\HH|)$ samples, which is a wider 
class than those of the previous results \cite{BCD05b,MR05d}.
It is noted that the essentially same upper bound\footnote{Strictly speaking, our bound is better than theirs up to a constant factor.} for CSI follows from the result of Ettinger et al.~\cite{EHK04}. However, their measurement is not
known to be a pretty good measurement.

We also present the following bounds of the sample complexity of TCS.
\begin{theorem}[Upper and Lower Bounds for TCS]\label{ULBforTCS}
Let $\HH$ be any set of candidate subgroups of a finite group.
Then, the sample complexity of TCS for $\HH$ is at most 
$O\left(\frac{\log|\HH|}{\log{\min_{H\in\HH}}|H|}\right)$.
If $|H|$ is a prime for every $H\in\HH$, the sample complexity is at least 
$\Omega\left(\frac{\log|\HH|}{\log{\max_{H\in\HH}}|H|}\right)$.
\end{theorem}

Summarizing these bounds, we obtain the following tight bounds for a class of CSI and TCS including several important instances such as $\HH_\mathrm{SDP}$ and $\HH_\mathrm{Sym}$.
\begin{corollary}\label{summary}
Let $\HH$ be any set of candidate subgroups of a finite group satisfying that 
$|H|=p$ for every $H\in\HH$, where $p$ is a prime.
Then, the sample complexity of CSI and TCS for $\HH$ is 
$\Theta\left(\frac{\log|\HH|}{\log{p}}\right)$.
\end{corollary}
This theorem implies that the decision version is as hard as the corresponding identification version in view of the sample complexity for this class.

We moreover apply our arguments to evaluation of
information-theoretic security of the quantum encryption schemes 
proposed by Kawachi et al.~\cite{KKNY05,KKNY06}.
They proposed two quantum encryption schemes: 
One is a single-bit encryption scheme, which has a 
computational security proof based on the worst-case hardness
of the decisional graph automorphism problem, and the other is a multi-bit 
encryption scheme, which has no security proof.
Since their schemes make use of quantum states quite similar to coset states 
over the symmetric group as the encryption keys and ciphertexts,
our proof techniques are applicable to the security evaluation of their 
schemes. We prove that the success probability of 
any computationally unbounded adversary distinguishing between 
any two ciphertexts is at most $\frac{1}{2}+2^{-\Omega(n)}$ 
in their $\log{m}$-bit encryption scheme with the security parameter $n$
if the adversary has only $o\left(\frac{n\log{n}}{m\log{m}}\right)$
encryption keys.

\section{Information-Theoretic Bounds}
In this section, we present the general bounds for CSI and TCS.
We first introduce basic notions and useful lemmas for 
our proofs in Section 2.1. 
We then give the general upper bounds for CSI and TCS in Section 2.2.
We also prove the general lower bounds for the sample complexity of CSI and TCS in Section 2.3.

\subsection{Basic Notions and Useful Lemmas}
Any quantum operations for extracting classical information from 
quantum states can be generally described by the positive operator-valued 
measure (POVM) \cite{NC00,Hay06}.
A POVM $M=\{M_i\}_{i\in S}$ associated with a set 
of outcomes $S$ is a set of Hermitian matrices satisfying 
that $M_i\geq 0\ (i\in S)$ and $\sum_{i\in S}M_i = I$. 
Then the probability of obtaining outcome $k\in S$ by the POVM $M$ 
from a quantum state $\rho$ is given by $\tr({M_k\rho})$.

The trace norm of a matrix $X\in\complex^{d\times d}$ is useful to 
estimate success probability of quantum state distinction for two 
states, and is defined as
$
 \tracenorm{X}= \max\limits_{\norm{Y}\leq 1}\langle Y,X  \rangle
              = \tr{\sqrt{X^\dag X}},
$
where $\norm{Y}$ is the $l_2$-norm of a matrix $Y$ and 
$\langle Y,X \rangle=\tr Y^\dag X$ is the matrix inner product.
It is well known that for any two quantum states $\rho_0$ and $\rho_1$ 
the average success probability of the optimal POVM 
distinguishing between two quantum states 
is equal to $\frac{1}{2}+\frac{1}{4}\tracenorm{\rho_0-\rho_1}$, i.e., 
$\frac{1}{2}\max_{M=\{M_0,M_1\}}(\tr{M_0\rho_0}+\tr{M_1\rho_1})=\frac{1}{2}+\frac{1}{4}\tracenorm{\rho_0-\rho_1}$.
See \cite{Bar97} for more details on the matrix analysis 
and \cite{NC00,Hay06} on basics of the quantum information theory.

We make use of the PGM in order to prove the general upper 
bound for CSI. The following lemma shown by Hayashi and Nagaoka~\cite{HN03}
is useful to estimate the error probability of the pretty good measurement.
(See also Lemma~4.5 in \cite{Hay06}.)
\begin{lemma}[Hayashi and Nagaoka \cite{HN03}]\label{HN03}
For any Hermitian matrices $S$ and $T$ satisfying that $I\geq S\geq 0$
and $T\geq 0$, it holds that 
$I-\sqrt{S+T}^{-1}S\sqrt{S+T}^{-1}\leq 2(I-S)+4T$,
where $\sqrt{S+T}^{-1}$ is the generalized inverse matrix of $\sqrt{S+T}$.
\end{lemma}

In our several proofs, we need to calculate the rank of a coset state.
The following lemma gives the estimation of the rank.
\begin{lemma}\label{rank}
For any coset state for a subgroup $H$ of a finite group $G$, 
it holds that $\rank{(\rho_H)}=\frac{|G|}{|H|}$.
\end{lemma}
\begin{proof}
Let $\ket{\psi}$ be a purification of $\rho_H$ described as 
$
 \ket{\psi}=\frac{1}{\sqrt{|G|}}\sum_{g\in G}\ket{g}_A\ket{f_H(g)}_B,
$
where $f_H$ is the given function in the definition of HSP. 
Tracing out the register $A$, we have
$\rank(\tr_A\ketbra{\psi}{\psi}) = |G/H|$. 
Since $\rank\left(\tr_A\ketbra{\psi}{\psi}\right) = \rank\left(\tr_B\ketbra{\psi}{\psi}\right)$, we obtain $\rank(\rho_H)=\frac{|G|}{|H|}$.
\end{proof}

\subsection{Lower Bounds}
We next prove the key theorem on lower bounds for CSI by a simple 
information-theoretic argument. This theorem generally 
gives the necessary number of identical samples of an unknown 
coset state for the identification.
\begin{theorem}\label{LBforCSI}
Let $\HH$ be any set of candidate subgroups of a finite group $G$.
Then, the sample complexity of CSI for $\HH$
is at least $\Omega\left(\frac{\log{|\HH|}}{\log\max_{H\in\HH}|H|}\right)$.
\end{theorem}
\begin{proof}
Let $M=\{M_H\}_{H\in\HH}$ be any POVM associated with 
$S_\HH$ using $k$ samples of the coset state.
By using the fact that 
$|\langle X,Y\rangle|\leq \norm{X}\tracenorm{Y}$
for any matrices $X,Y\in\complex^{d\times d}$,
the probability of $M$ obtaining correct outcome is upper bounded by
\begin{eqnarray*}
  \frac{1}{|\HH|}\sum_{H\in\HH}\tr M_H\rho_H^{\otimes k} 
&=& \frac{1}{|\HH|}\sum_{H\in\HH}\langle M_H,\rho_H^{\otimes k} \rangle\\
&\leq& \frac{1}{|\HH|}\sum_{H\in\HH} \Norm{\rho_H^{\otimes k}}\tracenorm{M_H}
= \frac{1}{|\HH|}\sum_{H\in\HH}\Norm{\rho_H^{\otimes k}}\tr\left(\sqrt{M_H^\dag M_H}\right)
= \frac{1}{|\HH|}\sum_{H\in\HH}\Norm{\rho_H^{\otimes k}}\tr M_H\\
&\leq& \frac{1}{|\HH|}\max_{H\in\HH}\norm{\rho_H}^k \sum_{H\in\HH}\tr M_H
= \frac{1}{|\HH|}\max_{H\in\HH}\norm{\rho_H}^k \tr\left(\sum_{H\in\HH} M_H\right)
= \frac{(\max_{H\in\HH}\norm{\rho_H} |G|)^k}{|\HH|}.
\end{eqnarray*}
Thus, the success probability of 
any quantum algorithm that solves CSI with $k$ coset states 
is upper bounded by 
$\frac{\left(\max_{H\in\HH}\Norm{\rho_{H}}|G|\right)^k}{|\HH|}$.
Since the coset state $\rho_{H}=\frac{1}{|G/H|}
\sum_{g\in G/H}\ketbra{gH}{gH}$ for any subgroup $H$
is a uniform summation of the matrices $\ketbra{gH}{gH}$ orthogonal 
to each other, 
we obtain $\Norm{\rho_{H}}=1/\rank(\rho_{H})$.
It follows that $\Norm{\rho_{H}}=|H|/|G|$ by Lemma~\ref{rank}.
The success probability 
is thus at most $\frac{\left(\max_{H\in\HH}|H|\right)^k}{|\HH|}$, 
which implies that any quantum algorithm that solves CSI for $\HH$ 
requires $\Omega\left(\frac{\log{|\HH|}}{\log{\max_{H\in\HH}|H|}}\right)$
coset states in order to attain constant success probability.
\end{proof}

As mentioned in Section 1, we do not have to identify a hidden subgroup to solve TCS. Thus, we cannot expect the same technique as the proof of the lower bound for CSI to work for that of TCS. We give another proof technique to obtain the lower bound for TCS.
\begin{theorem}\label{LBforTCS}
Let $\HH$ be any set of candidate subgroups of a finite group $G$. The sample complexity of TCS for $\HH$ is at least $\Omega\left(\frac{\log{|\HH|}}{\log\left(\max_{H\in\HH}|H|\right)}\right)$ if $|H|$ is a prime for every $H\in\HH$.
\end{theorem}
\begin{proof}
We first show that the success probability of solving TCS for $\HH$ 
is upper bounded by that of identification for certain two quantum states.
Let $M=\{M_0,M_1\}$ be any POVM associated with $\{\{id\},\HH\}$. The success 
probability of $M$ is given by 
$
 \min\{\tr M_0(I/|G|)^{\otimes k}, \min_{\rho_H\in S_\HH}\{\tr M_1\rho_H^{\otimes k}\} \}.
$
Also, it holds by the linearity of the trace and the POVM that
$
 \tr M_1\left(\frac{1}{|\HH|}\sum_{\rho_H\in S_\HH}\rho_H^{\otimes k}\right)
=\frac{1}{|\HH|}\sum_{\rho_H\in S_\HH}\tr M_1\rho_H^{\otimes k}
\geq \min_{\rho_H\in S_\HH} \tr M_1\rho_H^{\otimes k},
$
Thus, the success probability is at most
$
 \min\{\tr M_0(I/|G|)^{\otimes k}, \frac{1}{|\HH|}\sum_{\rho_H\in S_\HH}\tr M_1\rho_H^{\otimes k}\}.
$
This is equal to the success probability of the identification for 
$(I/|G|)^{\otimes k}$ and 
$\frac{1}{|\HH|}\sum_{\rho_H\in S_\HH}\rho_H^{\otimes k}$.

Note that we cannot apply the argument of Theorem~\ref{LBforCSI} to the identification.
Instead, we directly evaluate an upper bound of the trace norm of the matrix 
$X = \frac{1}{|\HH|}\sum_{\rho_H\in S_\HH}\rho_H^{\otimes k}
 - (I/|G|)^{\otimes k}$.
Then the success probability of the identification is at most
$\frac{1}{2}+\frac{1}{4}\tracenorm{X}$ by the property of the trace norm.
Na{\"{\i}}vely expanding $X$,
we obtain by the triangle inequality
\begin{eqnarray*}
\tracenorm{X} 
&=& \Tracenorm{\frac{1}{|\HH|}
   \sum_{H\in \HH}
   \frac{1}{|G|^k}\sum_{g_1,...,g_k\in G}
   \left(
   \sum_{h_1,...,h_k\in H}
   \ketbra{g_1,...,g_k}{g_1h_1,...,g_kh_k}
   -\ketbra{g_1,...,g_k}{g_1,...,g_k}
   \right)}\\
&=& \Tracenorm{\frac{1}{|\HH|}
   \frac{1}{|G|^k}\sum_{g_1,...,g_k\in G}
   \left(\sum_{H\in \HH}
   \sum_{\stackrel{h_1,...,h_k\in H}{(h_1,...,h_k)\neq(id,...,id)}}
   \ketbra{g_1,...,g_k}{g_1h_1,...,g_kh_k}
   \right)}\\
&\leq& 
   \frac{1}{|\HH||G|^k}
   \sum_{g_1,...,g_k\in G}
   \Norm{\ket{g_1,...,g_k}}
   \Norm{
   \sum_{H\in\HH}\sum_{
   \stackrel{h_1,...,h_k\in H}{(h_1,...,h_k)\neq(id,...,id)}
   }\bra{g_1h_1,...,g_kh_k}}\\
&=&
   \frac{1}{|\HH|}
   \sqrt{\left(\sum_{H,H'\in\HH}|H\cap H'|^k\right) - |\HH|^2}
= 
   \sqrt{\frac{1}{|\HH|^2}\left(\sum_{H\in\HH}|H|^k + \sum_{H\neq H'}|H\cap H'|^k \right)-1}\\
&\leq&
   \sqrt{
   \frac{\max_{H\in\HH}|H|^k}{|\HH|}
   }.
\end{eqnarray*}
In the last inequality, we use the fact that $|H\cap H'|=1$ for any distinct 
$H$ and $H'$, which follows from the prime order of the subgroups.

In order to have this trace norm larger than some positive constant,
$k$ must be 
$\Omega\left(\frac{\log{|\HH|}}{\log\left(\max_{H\in\HH}|H|\right)}\right)$.
Thus
$\Omega\left(\frac{\log{|\HH|}}{\log\left(\max_{H\in\HH}|H|\right)}\right)$
samples are necessary for constant advantage.
\end{proof}

\subsection{Upper Bounds}
We present general upper bounds for CSI and TCS in this section. 
First, we prove the upper bound for CSI by using the PGM for $S_\HH$. In this proof, we make use of Lemma~\ref{HN03} to estimate 
the error probability of the PGM. 
\begin{theorem}\label{UBforCSI}
Let $\HH$ be any set of candidate subgroups of a finite group $G$.
Then, the sample complexity of CSI for $\HH$ is at most
$O\left(\frac{\log|\HH|}{\log\min_{H\neq H'\in\HH}(|H|/|H\cap H'|)}\right)$.
\end{theorem}
\begin{proof}
Let $P_H$ be the projection onto the space spanned by $\supp(\rho_H)$ 
for $H\in\HH$.
We consider the pretty good measurement 
$M=\{\Sigma^{-1/2}P_H\Sigma^{-1/2}\}_{H\in\HH}$ for $S_\HH$, where 
$\Sigma=\sum_{H\in\HH}P_H$.
Let $\gamma_{H,H'}=|\{(h,h')\in H\times H': hh'=id\}|=|H\cap H'|$
for $H,H'\in\HH$. We now prove that the error probability of $M$
is at most $4\sum_{H'\neq H}\frac{(\gamma_{H,H'})^k}{|H'|^k}$
if the given state is $\rho_H$.

Since we have 
\begin{eqnarray*}
\tr\rho_H\rho_{H'}
= \frac{1}{|G|^2} \sum_{g,g'\in G} \sum_{h \in H, h' \in H' } 
\tr \ketbra{g}{gh}\ketbra{g'}{g'h'}
= \frac{1}{|G|^2} \sum_{g \in G} \sum_{h \in H, h' \in H' } 
\tr \ketbra{g}{gh h'}
= \frac{1}{|G|^2} \sum_{g \in G} \sum_{\stackrel{h \in H, h' \in H'}{hh'=id}}
1
= \frac{\gamma_{H,H'}}{|G|},
\end{eqnarray*}
it follows that 
$
 \tr P_H\rho_{H'} 
=\frac{\gamma_{H,H'}}{|G|}\frac{|G|}{|H'|}
=\frac{\gamma_{H,H'}}{|H'|}.
$
Setting $S=P_H^{\otimes k}$ and $T=\sum_{H'\neq H}P_{H'}^{\otimes k}$
in Lemma~\ref{HN03}, if the given state is $\rho_H$, the error probability 
of $M$ is 
\begin{eqnarray*}
\tr(I- \Sigma^{-1/2} P_H^{\otimes k} \Sigma^{-1/2})\rho_H^{\otimes k}
\leq 2\tr(I- P_H^{\otimes k})\rho_H^{\otimes k}
   + 4\tr\left(\sum_{H'\neq H} P_{H'}^{\otimes k}\right)\rho_H^{\otimes k} 
= 4 \sum_{H'\neq H}(\tr P_{H'}\rho_H)^{k}
= 4\sum_{H'\neq H}\frac{(\gamma_{H,H'})^{k}}{|H'|^k}.
\end{eqnarray*}

We can easily obtain the upper bound of the error probability from the above
estimation. Since we have
\[
 4\max_{H\in\HH}\sum_{H'\neq H}\frac{(\gamma_{H,H'})^{k}}{|H'|^k}
\leq 4|\HH|\max_{H\neq H'\in\HH}\left(\frac{|H\cap H'|}{|H|}\right)^k,
\]
the error probability of $M$ is at most 
$4|\HH|\max_{H\neq H'\in\HH}\left(\frac{|H\cap H'|}{|H|}\right)^k$,
which implies that 
$O\left(\frac{\log|\HH|}{\log\min_{H\neq H'\in\HH}(|H|/|H\cap H'|)}\right)$ samples of coset states are sufficient for constant success probability.
\end{proof}

Next, we present the general upper bound for TCS as follows. 
This upper bound can be attained by a simple two-valued POVM.
\begin{theorem}\label{UBforTCS}
Let $\HH$ be any set of candidate subgroups of a finite group $G$. Then 
the sample complexity of TCS for $\HH$ is at most $O\left(\frac{\log|\HH|}{\log\min_{H\in\HH}|H|}\right)$.
\end{theorem}
\begin{proof}
We consider a projection $T$ onto the space spanned by 
$\bigcup_{H\in\HH}\supp(\rho_H^{\otimes k})$.
It obviously holds that $\tr T \rho_H^{\otimes k} = 1$ for every $H\in\HH$.
On the other hand, the error probability is given by 
$\tr T(I/|G|)^{\otimes k}$. Then we have 
$
 \tr T(I/|G|)^{\otimes k} = \frac{\rank(T)}{|G|^k} \leq \frac{\sum_{H\in\HH}\rank(\rho_H)^k}{|G|^k}.
$
Since $\rank(\rho_H) = |G|/|H|$ by Lemma~\ref{rank},
we obtain 
$\frac{\sum_{H\in\HH}\rank(\rho_H)^k}{|G|^k}
=\frac{\sum_{H\in\HH}(|G|/|H|)^k}{|G|^k}
\leq \frac{|\HH|}{\min_{H\in\HH} |H|^k}$.
This implies that at most 
$O\left(\frac{\log|\HH|}{\log\min_{H\in\HH}|H|}\right)$ samples of
coset states are sufficient for constant advantage.
\end{proof}

\section{Security Evaluation of Quantum Encryption Schemes}
Our arguments are applicable not only to bounds for HSP but also to 
security evaluation of quantum cryptographic schemes.
In this section, we apply our arguments to evaluation of the 
information-theoretic security of the quantum encryption schemes 
proposed in \cite{KKNY05,KKNY06}. As mentioned in Section 1,
they proposed single-bit and multi-bit quantum encryption schemes.
While they gave the complexity-theoretic security to 
the single-bit scheme under the assumption of the worst-case hardness 
of the decisional graph automorphism problem, the multi-bit one has 
no security proof. Also, they have already proven in \cite{KKNY06} that any 
computationally unbounded quantum algorithm 
cannot solve a certain quantum state distinction problem that underlies 
the single-bit scheme with few samples by reducing the solvability 
of their distinction problem to the result of \cite{HMRRS06}.
On the other hand, the security of their encryption schemes, 
as well as the underlying problem for their multi-bit scheme, are not 
evaluated yet from a viewpoint of the quantum information theory. 

Their schemes make use of certain quantum states 
for their encryption keys and ciphertexts.
We now call these quantum states {\em encryption-key states\/} and 
{\em cipherstates\/}, respectively. 
Since their multi-bit encryption scheme contains 
the single-bit one as a special case if we ignore its efficiency and 
complexity-theoretic security, 
we only discuss their multi-bit scheme in this paper.

We now describe their multi-bit encryption scheme in detail.
Assume that the message length parameter $m$ divides the security 
parameter $n$, where $m\in\{2,...,n\}$.
Let $\KK_n^m=\{h: h=(a_1\cdots a_m)\cdots(a_{n-m+1}\cdots a_n),\ a_i\in\{1,...,n\},\ a_i \neq a_j\ (i\neq j)\}\subset S_n$,
i.e., a set of the permutations composed of $n/m$ disjoint 
cyclic permutations, which is used for the decryption key. 
In this scheme, we exploit the following quantum state for a message $s$:
$
 \rho_h^{(s)} = \frac{1}{mn!}\sum_{g\in S_n}
\left(\sum_{k=0}^{m-1}\omega_m^{ks}\ket{gh^k}\right)
\left(\sum_{l=0}^{m-1}\omega_m^{-ls}\bra{gh^l}\right),
$
where $\omega_m = e^{2\pi i/m}$ and $h\in\KK_n^m$. 
Note that $\rho_h^{(0)}$ is the coset state for 
the hidden subgroup $\{id,h,...,h^{m-1}\}$.

We now refer to as $(n,m)$-QES their multi-bit encryption scheme with 
the security parameter $n$ and the message length parameter $m$. 
The protocol of $(n,m)$-QES is summarized as follows.
\begin{description}
\magicwand
\item[Protocol: $(n,m)$-QES]
\item[{\rm (1)}] The receiver Bob chooses his decryption key $h$
uniformly at random from $\KK_n^m$ 
and generates the encryption-key states $\sigma_h=(\rho_h^{(0)},...,\rho_h^{(m-1)})$.
\item[{\rm (2)}] The sender Alice requests the encryption-key state $\sigma_h$ 
to Bob. She picks $\rho_h^{(s)}$ up from $\sigma_h$ as the cipherstate 
corresponding to her classical message $s\in\{0,...,m-1\}$
 and then sends it to him.
\item[{\rm (3)}] Bob decrypts her cipherstate $\rho_h^{(s)}$ with his decryption key $h$.
\end{description}

We assume the same adversary model except for Eve's computational power 
as the original ones in \cite{KKNY05,KKNY06}. 
Note that the eavesdropper Eve can also request the same 
encryption-key states to Bob as one of senders.
Eve in advance requests the encryption-key states to Bob. 
When Alice sends to Bob her cipherstate that Eve wants to eavesdrop,
Eve picks up Alice's cipherstate and then 
tries to extract Alice's message from the cipherstate with the 
encryption-key states by 
computationally unbounded quantum computer, i.e., Eve can apply an
arbitrary POVM over the cipherstates and encryption-key states 
to extract Alice's message.

We consider a stronger security notion such that 
Eve cannot distinguish between even two candidates, i.e., 
she cannot find a non-negligible gap between 
$\tr M_1(\rho_h^{(s)}\otimes\sigma_h^{\otimes k})$ and 
$\tr M_1(\rho_h^{(s')}\otimes\sigma_h^{\otimes k})$ 
even by the optimal POVM $M=\{M_0,M_1\}$ 
when Bob chooses $h$ uniformly at random.
This notion naturally extends the computational 
indistinguishability of encryptions, which is
the standard security notion in the modern cryptography~\cite{Gol01}, 
to the information-theoretic one.

Since the gap is at most $\frac{1}{2}\tracenorm{\frac{1}{|\KK_n^m|}\sum_{h\in\KK_n^m}\rho_h^{(s)}\otimes\sigma_h^{\otimes k}-\rho_h^{(s')}\otimes\sigma_h^{\otimes k}}$, this notion can be formalized by the trace norm between them.
Then, we say that the cipherstates are {\em information-theoretically 
indistinguishable within $k$ encryption-key states\/} 
if $\tracenorm{\frac{1}{|\KK_n^m|}\sum_{h\in\KK_n^m}\rho_h^{(s)}\otimes\sigma_h^{\otimes k}-\rho_h^{(s')}\otimes\sigma_h^{\otimes k}}=2^{-\Omega(n)}$.

For this security notion, we can obtain the following theorem 
by our information-theoretic arguments.
The proof is almost straightforward by Theorem~\ref{LBforTCS}.
\begin{theorem}\label{IT-security}
The cipherstates of $(n,m)$-QES are information-theoretically 
indistinguishable within $o\left(\frac{n\log{n}}{m\log{m}}\right)$ 
encryption-key states.
\end{theorem}
\begin{proof}
Let $l_s =  \Tracenorm{\frac{1}{|\KK_n^m|}\sum_{h\in\KK_n^m}\rho_h^{(s)}\otimes\sigma_h^{\otimes k}-(I/n!)^{\otimes mk+1}}$.
Then the trace norm between two state sequences given in the definition 
of the information-theoretic indistinguishability
is at most $l_s + l_{s'}$ by the triangle inequality.
Since the trace norm is invariant under unitary 
transformations, we can show that $l_s+l_{s'}=2l_0$
by taking appropriate unitary operators.
Then we can prove that $l_0\leq \sqrt{m^{mk+1}/|\KK_n^m|}$ by the argument of Theorem~\ref{LBforTCS}. Since we have 
$|\KK_n^m|\approx \frac{m^{1/2}n^{n-n/m}}{e^{n-n/m}}$ by 
the standard counting method and the Stirling approximation,
the trace norm is at most $2^{-\Omega(n)}$ 
if $k=o\left(\frac{n\log{n}}{m\log{m}}\right)$.
\end{proof}\\
For example, when we set $m=n^\varepsilon$ for any constant $0<\varepsilon<1$,
we obtain the $\varepsilon\log{n}$-bit encryption scheme whose cipherstates are
 information-theoretically indistinguishable within $o(n^{1-\varepsilon})$ 
encryption-key states.

\section{Concluding Remarks}
In this paper, we have shown general bounds for CSI and TCS, and 
an application to the security evaluation of the quantum encryption schemes. We believe such an information-theoretic approach will help constructions of efficient quantum algorithms for non-Abelian HSPs as in the case of \cite{BCD05b}.
After our preliminary version of this paper, Harrow and Winter followed our approach to prove the existence of a quantum measurement for identifying general quantum states and lower bounds of samples for the identification \cite{HW06}. Their results generalize and improve our bounds for CSI.

$\mbox{}$\\
\noindent{\bf Acknowledgements.}
The authors would like to thank Fran\c{c}ois Le Gall, Cristopher Moore, Christopher Portmann and Tomoyuki Yamakami for helpful discussions and comments.
MH is grateful to Hiroshi Imai with the ERATO-SORST QCI project for support.
AK was supported by the Ministry of Education, Science, Sports and Culture, 
Grant-in-Aid for Young Scientists (B) No.17700007, 2005 and for Scientific 
Research on Priority Areas No.16092206.


\end{document}